\documentclass[a4paper,12pt]{article}
\pdfoutput=1 

\usepackage{jheppub} 

\usepackage[T1]{fontenc} 

\usepackage[multiple]{footmisc}

\usepackage{graphicx}
\usepackage{amssymb, amsmath,mathtools,amsthm}
\usepackage{hyperref}
\hypersetup{
    colorlinks=true,
    linkcolor=black,          
    citecolor=black,        
    filecolor=black,      
    urlcolor=black          
}

\DeclareMathOperator{\Tr}{Tr}

\newcommand{\beq}{\begin{equation}}
\newcommand{\eeq}{\end{equation}}

\newcommand{\sdiff}{SDiff($S^2$)}
\newcommand{\svect}{SVect($S^2$)}
\newcommand{\svectstar}{SVect($S^2$)$^*$}

\def\div{{\rm div}}

\title{\boldmath ${\rm SDiff}(S^2)$ and the orbit method}

\author{Robert Penna}
\affiliation{Institute for Advanced Study, Princeton, NJ 08540, USA}
\emailAdd{rpenna@ias.edu}

\abstract{The group of area preserving diffeomorphisms of the two sphere, ${\rm SDiff}(S^2)$, is one of the simplest  examples of an infinite dimensional Lie group.  It plays a key role in incompressible hydrodynamics and it recently appeared in general relativity as a subgroup of two closely related, newly defined symmetry groups.  We investigate its representation theory using the method of coadjoint orbits.  We describe the Casimir functions and the Cartan algebra.  Then we evaluate the trace of a simple ${\rm SDiff}(S^2)$ operator using the Atiyah-Bott fixed point formula.  The trace is divergent but we show that it has well-defined truncations related to the structure of ${\rm SDiff}(S^2)$.  Finally, we relate our results back to the recent appearances of ${\rm SDiff}(S^2)$ in black hole physics.}

\begin{document} 
\maketitle
\flushbottom

\section{Introduction}
\label{sec:intro}

The set of area preserving diffeomorphisms of the two-sphere, \sdiff, is a basic example of an infinite dimensional Lie group.  It is the configuration space and the symmetry group of the incompressible Euler equations for fluid flow on the sphere \cite{arnold1966geometrie,arnold1969hamiltonian,arnold1999topological,khesin2008geometry,marsden2013introduction}.  It may have a role to play in black hole physics, where it is a subgroup of two recently introduced symmetry groups.  One acts on the black hole event horizon \cite{Donnay:2015abr,Donnay:2016ejv,Chandrasekaran:2018aop} and the other acts on the asymptotic boundary of asymptotically flat spacetimes \cite{Campiglia:2015yka,Compere:2018ylh,Flanagan:2019vbl}.  The possibility of connections to black hole physics was the original motivation for the present work.

Elements of \sdiff\ are smooth invertible maps from the sphere to itself that preserve the area form.  Composition of two maps,  $\eta_1 \circ \eta_2$, gives the group structure.   The tangent space of \sdiff\ at the identity map is the space of divergence-free vector fields on the sphere, \svect.  The Lie bracket of two divergence-free vector fields is minus the usual vector field commutator.  In a coordinate chart, the usual vector field commutator is
\beq
[v_1,v_2]^j = v_1^i \frac{\partial v_2^j}{\partial x^i}  - v_2^i \frac{\partial v_1^j}{\partial x^i}.
\eeq

In the context of incompressible fluid dynamics on the sphere, the maps from Lagrangian coordinates (a choice of fluid frame) to Eulerian coordinates (a choice of space fixed frame) are elements of \sdiff.  The fluid's instantaneous velocity field is an element of \svect.   The incompressible Euler equations have an infinite dimensional ``particle relabeling symmetry,'' which comes from an action of \sdiff\ on phase space.  This symmetry gives an infinite number of conservation laws, which are equivalent to Kelvin's circulation theorem.  In the context of hole hole physics, the analogous conservation laws might be related to ``superrotation'' charge conservation \cite{Barnich:2011ct,Barnich:2011mi,Pasterski:2015tva,Penna:2015gza,Penna:2017vms}.

A simple ``toy model'' for \sdiff\ is $SO(3)$, the rotation group of three dimensional Euclidean space.  Rotations act on the sphere as area preserving diffeomorphisms, so \sdiff\ represents a kind of infinite dimensional enhancement of $SO(3)$.  The relationship between \sdiff\  and incompressible fluid dynamics has a close analogue in the relationship between $SO(3)$ and rigid body dynamics \cite{arnold1966geometrie,arnold1969hamiltonian,arnold1999topological,khesin2008geometry,marsden2013introduction}.

So far we have been describing applications from classical physics, but $SO(3)$ also governs the quantum mechanics of a spinning particle. It is natural to ask if \sdiff\ has an analogous role to play in quantum field theory, perhaps for some generalization of the spinning particle.  It would be especially interesting if such a quantum system could be related back to the appearances of \sdiff\ in black hole physics.

We do not construct such a quantum system in this paper, but instead describe some of the properties such a system would have.  We do this by investigating the representation theory of \sdiff\ using the method of coadjoint orbits.  The orbit method is reviewed in \cite{vergne1983representations,witten1988coadjoint,kirillov2004lectures}.  We do not assume any familiarity but develop the theory as we go along, using the simpler case of $SO(3)$ and the spinning particle as a guide.  

A spinning particle's quantum states can be labeled by the eigenvalues of two operators, $\hat{J}^2$ and $\hat{J}_z$,  the square of the total angular momentum and the angular momentum along the $z$-axis.  In Section \ref{sec:labels}, we describe the analogue of these operators for \sdiff.  In Section \ref{sec:traces}, we use the orbit method to study the traces of these operators.  Our main result is formula \eqref{eq:tracexi} for the trace of an \sdiff\ operator.   In Section \ref{sec:discuss}, we relate our results back to the recent appearances of \sdiff\ in black hole physics.

\sdiff\ is a well studied group.  Applications to incompressible fluid dynamics have been a major source of motivation  \cite{arnold1966geometrie,arnold1969hamiltonian,arnold1999topological,khesin2008geometry,marsden2013introduction}.  Izosimov, Khesin, and Mousavi \cite{izosimov2016coadjoint} recently gave a complete classification of generic coadjoint orbits.  The representation theory is less well understood.  As far as we know, our formula for the trace \eqref{eq:tracexi} is new, as is the context relating our results to black hole physics.

\section{Labeling the states}
\label{sec:labels}

The states of a quantum spinning particle can be labeled by the eigenvalues of two operators, $\hat{J}^2$ and $\hat{J}_z$.  The goal of the present section is to find the analogue of these operators for \sdiff.  We use the method of coadjoint orbits.  We develop the theory as we go along, using $SO(3)$ and the spinning particle as a guide.

The angular momentum of a classical spinning particle is a three-vector, $\vec{J}$.  The space of states of a classical spinning particle with fixed $J^2$ is a two sphere, $S^2_{J^2} \subset \mathbb{R}^3$, with radius $|J|$.  Classically, any angular momentum vector on $S^2_{J^2}$ is allowed.  Quantum mechanically, the angular momentum is an operator and $\hat{J}_x$, $\hat{J}_y$, and $\hat{J}_z$ are not all independent.  A quantum state is uniquely fixed by the eigenvalues of $\hat{J}^2$ and $\hat{J}_z$ alone.  This reflects the fact that the quantum operators $\hat{J}_x$, $\hat{J}_y$, and $\hat{J}_z$ do not commute.  The nonvanishing commutators are
\beq\label{eq:liealg}
[\hat{J}_x,\hat{J}_y] = i\hat{J}_z, \quad
[\hat{J}_y,\hat{J}_z] = i\hat{J}_x, \quad
[\hat{J}_z,\hat{J}_x] = i\hat{J}_y,
\eeq
which is the Lie algebra $\mathfrak{so}(3)$.  Quantum states are labeled by $\hat{J}^2$ and by a set of mutually commuting generators of the Lie algebra.  $\hat{J}_z$ commutes with itself but not with either of the other two generators, so the eigenvalues of $\hat{J}^2$ and $\hat{J}_z$ suffice to label the states.  

Now we want to extend this discussion to \sdiff.  First we need the analogues of $\vec{J}\in \mathbb{R}^3$ and $S^2_{J^2}$.  The spinning particle's classical angular momentum vector, $\vec{J}$, acts on the spinning particle's phase space as an infinitesimal rotation generator.  The infinitesimal generators of \sdiff\ transformations are divergence-free vector fields.  So the analogue of $\mathbb{R}^3$ is the infinite dimensional vector space \svect.

Acting with $SO(3)$  on a fixed angular momentum vector $\vec{J}\in \mathbb{R}^3$ sweeps out the two-sphere,
\beq
S^2_{J^2} = \{g\cdot \vec{J} | g\in SO(3)\} \subset \mathbb{R}^3.
\eeq
The action of  $SO(3)$ on $\mathbb{R}^3$ is the adjoint action and $S^2_{J^2}$ is an adjoint orbit.
The analogues of the two-spheres for \sdiff\ are the orbits of the adjoint action of \sdiff\ on \svect.  Classifying the adjoint orbits gives the classical analogues of the $\hat{J}^2$ operator.  This classification has been known heuristically for a long time but a rigorous proof was only given recently \cite{izosimov2016coadjoint}.  We turn to a description of these results.

\subsection{Coadjoint orbits and Casimir functions}

The adjoint action of \sdiff\ on \svect\ is defined by first considering the action of \sdiff\ on itself by conjugation,
\beq\label{eq:conjug}
\eta'\rightarrow \eta \circ \eta' \circ \eta^{-1},
\eeq
and then studying the limit where $\eta'$ is close to the identity.  This means we can expand 
\beq
\eta' = {\rm id.} + t \xi + O(t^2),
\eeq
where $\xi$ is a divergence-free vector field.  Plugging into \eqref{eq:conjug} and differentiating with respect to $t$ at $t=0$ defines the adjoint action of $\eta$ on $\xi$.  

This can be understood more simply using coordinates.  Set $x\equiv \eta(X)$ and think of $\eta$ as the change of coordinates $X^i\rightarrow x^i$.    Then the adjoint action sends $\xi^i(X) \rightarrow \frac{\partial x^i}{\partial X^j} \xi^j(X)$.  Our goal is to describe the orbits of this action.

We can make things easier by switching to the dual vector space \svectstar.   The dual of a vector field is one-form.  To define \svectstar, we need an analogue of the divergence-free condition for the one-forms.  A one-form, $u\in \Omega^1(S^2)$, gives a linear map on \svect\ as follows:
\beq\label{eq:linear}
u(\xi) \equiv \int_{S^2} u \wedge \iota_\xi \mu,
\eeq
where $\iota_\xi \mu$ is the interior product with the area form, $\mu$, on the two-sphere.  Now if $u=df$ is exact, then
\beq\label{eq:df}
df(\xi) =  \int_{S^2} df \wedge \iota_\xi \mu =  \int_{S^2} d(f \wedge \iota_\xi \mu) = 0,
\eeq
where we used Cartan's identity and $\div\xi=0$ to set $d\iota_\xi \mu = L_\xi \mu - \iota_\xi d\mu = 0$.  Equation \eqref{eq:df} shows that exact forms annihilate \svect.   To define \svectstar, we should therefore identify two one-forms $u$ and $u'$ if they differ by an exact form.  This is the analogue of the divergence-free condition.  It gives the coset construction ${\rm SVect}(S^2)^* = \Omega^1(S^2)/d\Omega^0(S^2)$.  The effect of switching from \svect\ to its dual \svectstar\ has been to replace the constraint $\div\xi=0$ with the gauge symmetry $u\rightarrow u+df$.  

In coordinates, the linear map \eqref{eq:linear} is
\beq
u(\xi) = \int_{S^2} u_i \xi^i \mu.
\eeq
Recall that the adjoint action sends $\xi^i \rightarrow \frac{\partial x^i}{\partial X^j} \xi^j$.  The dual of the adjoint action is evidently $u_j\rightarrow u_i\frac{\partial x^i}{\partial X^j}$.  This is an antirepresentation, meaning the action of the product $\eta_1 \eta_2$ is the product of the actions of $\eta_2$ and $\eta_1$ in reverse order.  To get an ordinary representation, the coadjoint action is defined as $u_j\rightarrow u_i\frac{\partial X^i}{\partial x^j}$.  In coordinate free notation, the coadjoint action is the pull back of $u$ by the inverse map,
\beq
\eta \cdot u \equiv (\eta^{-1})^* u.
\eeq
Using the two-sphere metric to identify upper and lower indices gives an isomorphism between the adjoint and coadjoint actions.

Recall that the space of classical states of a spinning particle with angular momentum $|J|$ is the two-sphere, $S^2_{J^2}$, and the spheres are labeled by a single parameter, $J^2$.  The analogue of the sphere for \sdiff\ is the coadjoint orbit
\beq\label{eq:orbits}
\mathcal{O}_u \equiv \{\eta \cdot u | \eta \in {\rm SDiff}(S^2)\}.
\eeq
An infinite number of parameters are needed to distinguish the coadjoint orbits.   Each parameter is a real-valued functional on \svectstar.  A functional on \svectstar\ is the same thing as a gauge invariant functional on $\Omega^1(S^2)$.  Such a functional is constant on coadjoint orbits if it is coordinate independent.  An infinite number of such functionals is provided by
\beq\label{eq:Ik}
I_k(u) \equiv \int_{S^2} \left(\frac{du}{\mu}\right)^k \mu, \quad k=2, 3, \dots.
\eeq
The $I_k(u)$ are manifestly coordinate independent and gauge invariant, $u\rightarrow u+df$.  In the context of incompressible fluid dynamics, $u$ is the fluid's instantaneous (co)velocity and $I_k(u)$ is the $k$'th moment of the vorticity function, $du/\mu$.

The vorticity moments, $I_k(u)$, $k=2,3,\dots$, are the analogue of $J^2$.  They provide a set of labels for the coadjoint orbits.   They are not, however, a complete set of labels.  Roughly speaking, to completely distinguish generic coadjoint orbits one needs to keep track of the vorticity moments on each component of the level set of the vorticity function.  For the precise statement see \cite{izosimov2016coadjoint}.

There is another characterization of the $I_k(u)$ that is worth mentioning.  As befits a space of classical states, \svectstar\ has a Poisson bracket, the Lie-Poisson bracket,
\beq\label{eq:liepoisson}
\{F,G\}(u)  \equiv u \left(\left[\frac{\delta F}{\delta u},\frac{\delta G}{\delta u}\right] \right),
\eeq
where $F(u)$ and $G(u)$ are functionals on \svectstar.   Consider $\{F,I_k\}$, for $F$ arbitrary.  The Lie bracket, $[\delta F/\delta u, \delta I_k/\delta u]$, represents the action of the vector field $\delta F/\delta u$ on the vector field $\delta I_k/\delta u$ by infinitesimal coordinate transformations.  This boils down to computing
\beq
\delta I_k 
	= \int_{S^2} k\left(\frac{du}{\mu}\right)^{k-1} d\delta u
	= -\int_{S^2} k d\left(\frac{du}{\mu}\right)^{k-1} \wedge \delta u
	= \int_{S^2} \langle *d\left(\frac{du}{\mu}\right)^{k-1}k, \delta u \rangle \mu,
\eeq
which gives $\delta I_k/\delta u  = *d(du/\mu)^{k-1}k$, which is a manifestly coordinate independent vector field.  We have thus shown that 
\beq
\{ F, I_k \} = 0,\quad k=2,3,\dots,
\eeq
for $F$ arbitrary.  In other words, the $I_k(u), k=2,3,\dots$, are classical Casimir functions of \sdiff. 

\subsection{$SO(3)$ revisited}

In the process of defining the vorticity moments, $I_k(u), k=2,3,\dots$, we were led to introduce the coadjoint action, coadjoint orbits, and the Lie-Poisson bracket.  Having established this machinery, it is perhaps worth revisiting the simpler example of $SO(3)$.  This will firm up the analogy between the vorticity moments and $J^2$.

The Lie algebra, $\mathfrak{so}(3)$, is the space of $3\times3$ antisymmetric matrices.  The adjoint action is $X\rightarrow gXg^{-1}, X\in \mathfrak{so}(3), g\in SO(3)$.  We identify $\mathfrak{so}(3)\approx \mathbb{R}^3$ using the map
\beq
\begin{pmatrix}
0	&	-X_3	&	X_2 \\
X_3 	&	0	&	-X_1	\\
-X_2	&	X_1	&	0
\end{pmatrix}
\leftrightarrow	(X_1, X_2, X_3).
\eeq
The matrix commutator becomes the vector cross product and the adjoint action becomes $\vec{J}\rightarrow g \vec{J}, \vec{J}\in \mathbb{R}^3$.  The adjoint orbits are the two-spheres, $S^2_{J^2}$.

As vector spaces, $\mathbb{R}^{3*}\approx \mathbb{R}^3$.  Vectors act as linear maps via 
\beq
\vec{J_1}(\vec{J_2})  = \vec{J_1}^T\vec{J_2}.
\eeq
Under the adjoint action, $\vec{J_2}\rightarrow g\vec{J_2}$.  So the dual of the adjoint action is $\vec{J_1}\rightarrow g^T \vec{J_1}$.  As before, this is an antirepresentation, and we get an ordinary representation by exchanging $g$ with $g^{-1}=g^T$ in the definition of the coadjoint action: $\vec{J_1}\rightarrow g \vec{J_1}$.  So again the adjoint and coadjoint actions are isomorphic\footnote{This is not always the case.  For example, the adjoint and coadjoint actions of the Virasoro group are not isomorphic.  In general, coadjoint orbits are more important than adjoint orbits.  Quantizing coadjoint orbits gives irreducible representations.}.  The coadjoint orbits are also two-spheres, $S^2_{J^2}$.

The Lie-Poisson bracket on $\mathbb{R}^3$ is
\beq
\{F,G\}(\vec{J}) = \vec{J}\cdot \left(\nabla F \times \nabla G \right).
\eeq
Following our earlier discussion, we anticipate $\{F,J^2\} = 0$, $F$ arbitrary.  Indeed, $\nabla J^2 = 2(J_1,J_2,J_3)=2\vec{J}$, so the cross product is orthogonal to $\vec{J}$.  This shows $J^2$ is a classical Casimir function for $SO(3)$.

\subsection{Cartan algebra}

Classically, any state on $S^2_{J^2}$ is allowed, because $J_x, J_y$, and $J_y$ are commuting observables.  Quantization turns $J_x, J_y,$ and $J_z$ into operators with commutators given by the $\mathfrak{so}(3)$ Lie algebra \eqref{eq:liealg}.  Quantum states are labeled by the eigenvalues of $\hat{J}^2$ and $\hat{J}_z$ alone.  The operator $\hat{J}_z$ commutes with itself but not with $\hat{J}_x$ or $\hat{J}_z$.  It spans a maximal commutating subalgebra, or Cartan algebra, of $\mathfrak{so}(3)$.  The goal of the present section is to find the analogue of $J_z$ for \sdiff. 

The definition of \sdiff\ requires a choice of area form, $\mu$.  So far $\mu$ has been arbitrary.  To keep things simple going forward, the area form is now the usual one, $\mu=\sin\theta d\theta \wedge d\varphi$.

The Lie algebra is \svect. To get a basis, let $Y_{\ell m}$ be a spherical harmonic function.  The vector field $\nabla Y_{\ell m}$ is curl free.   To get a divergence-free vector field, set
\beq\label{eq:basis}
\xi_{\ell m} \equiv n \times \nabla Y_{\ell m},
\eeq
where $n$ is the unit outward normal to the sphere.  Now $\nabla \cdot \xi_{\ell m} = 0$ follows from the identity $\nabla \cdot (A\times B) = (\nabla \times A) \cdot B - (\nabla \times B) \cdot A$.  In the language of differential forms, $\iota_{\xi_{\ell m}} \mu = -dY_{\ell m}.$  This formula makes it clear that the relationship between $\xi_{\ell m}$ and $Y_{\ell m}$ does not depend on the metric.  It only depends on the area form.  So it is invariant under \sdiff\ transformations. The $\xi_{\ell m}$ form a basis for \svect.  In coordinates,
\beq
\xi_{\ell m} = -(\sin \theta )^{-1} \partial_\varphi Y_{\ell m} \frac{\partial}{\partial\theta}
 + (\sin\theta)^{-1} \partial_\theta Y_{\ell m} \frac{\partial}{\partial \varphi}.
\eeq
It is not too hard to check that 
\beq\label{eq:comm}
[\xi_{\ell m},\xi_{\ell' m'}] = -n\times \nabla \{Y_{\ell m},Y_{\ell' m'}\},
\eeq
where
\beq\label{eq:Ypb}
\{Y_{\ell m},Y_{\ell' m'}\} = (\sin\theta)^{-1}\left(
\frac{\partial Y_{\ell m}}{\partial \varphi}    \frac{\partial Y_{\ell' m'}}{\partial \theta}
-\frac{\partial Y_{\ell m}}{\partial \theta}    \frac{\partial Y_{\ell' m'}}{\partial \varphi} \right).
\eeq
The $m=0$ harmonics commute: $[\xi_{\ell 0},\xi_{\ell' 0}] = 0$ for all $\ell \geq 1, \ell' \geq 1$.

The $\xi_{\ell m}$ can be thought of as coordinate functions on \svectstar, by setting
\beq\label{eq:coords}
\xi_{\ell m}(u) \equiv u(\xi_{\ell m}) 
= \int_{S^2} \xi_{\ell m}^i u_i\mu
=-\int_{S^2} Y_{\ell m} du.
\eeq
As such, we can take Poisson brackets using \eqref{eq:liepoisson},
\beq\label{eq:pbs}
\{\xi_{\ell m},\xi_{\ell ' m'} \}(u)
=u([\xi_{\ell m},\xi_{\ell' m'}])
=[\xi_{\ell m},\xi_{\ell' m'}](u).
\eeq
Quantization sends $\{ \xi_{\ell m},\xi_{\ell' m'} \}\rightarrow \frac{1}{i \hbar}[ \hat{\xi}_{\ell m}, \hat{\xi}_{\ell' m'} ]$.  We will not attempt to make sense of the rhs, but assume that it exists.  Equation \eqref{eq:pbs} implies the commutators of the quantum operators are the same thing as the \svect\ Lie algebra commutators.  In the quantum theory, the failure of $\hat{\xi}_{\ell m}$ and $\hat{\xi}_{\ell' m'}$ to commute means they are not independent observables.  However, the $m=0$ operators all commute. These operators are the analogue of $\hat{J}_z$.  

Table \ref{tab:table} summarizes the results of this section.

\begin{table}[ht]
\centering
\caption{}
\vspace{0.2cm}
\begin{tabular}{c c c}
\hline
					&	$SO(3)$			&	\sdiff\			\\
\hline					
Lie algebra			&	$\mathbb{R}^3$	&	\svect\			\\
dual Lie algebra		&	$\mathbb{R}^3$	&	$\Omega^1(S^2)/d\Omega^0(S^2)$		\\
classical state			&	$\vec{J}$			&	$u$								\\
coadjoint orbit			&	$S^2_{J^2}$		&	$\mathcal{O}_u$					\\
Casimir functions		&	$J^2$ 			& 	$I_k(\mu), \quad k=2,3,\dots$			\\
Cartan algebra			&	$J_z$			&	$\xi_{\ell 0}, \quad \ell	=1,2,\dots$	\\
physical model			&	spinning particle	&	incompressible fluid	
\end{tabular}
\label{tab:table}
\end{table}

\section{Traces}
\label{sec:traces}

The quantum states of a spinning particle are labeled by the eigenvalues of $\hat{J}^2$ and $\hat{J}_z$. It is easy to list the eigenvalues explicitly\footnote{We are sticking with $SO(3)$ rather than passing to its double cover $SU(2)$ which is why we do not consider the half-spin states.}: the eigenvalues of $\hat{J}^2$ are $\ell (\ell +1)$, $\ell=0,1,2,\dots$, and the eigenvalues of $\hat{J}_z$ are $m=-\ell,-\ell + 1 , \dots,\ell$.   

The situation for \sdiff\ is much more complicated.  Finding the eigenvalues of the $\hat{I}_k$ and $\hat{\xi}_{\ell 0}$ operators is beyond the scope of this paper.  A more tractable problem is to study the traces of these operators.  This is the goal of the present section.

Returning to $SO(3)$, fix $\ell$ and consider
\beq\label{eq:trace}
\Tr q^{\hat{J}_z} 
	= q^{-\ell} + q^{-\ell+1} + \dots + q^{\ell} 
	= \frac{q^{\ell+1}-q^{-\ell}}  {q-1},
\eeq
where $q$ is a formal complex variable.  The second equality perhaps deserves a comment.  We used $(q-1)^{-1}=-1-q-q^2-\dots$ near $q=0$.  We get two copies of this series, one for each term in the numerator, $q^{\ell+1}-q^{-\ell}$.  Taking their difference kills all but a finite number of terms.  This shows that the second equality in \eqref{eq:trace} is exact for $|q|<1$.  The same argument works near $q=\infty$, so \eqref{eq:trace} is exact for $|q|\neq 1$.  The rhs is singular at $q=1$ but the limit is well defined.  It equals $2\ell+1$, the number of quantum states with fixed $\ell$.

Computing the trace as in \eqref{eq:trace} requires the eigenvalues of $\hat{J}_z$, so we do not seem to be any better off.  However, there are independent definitions of the trace that only require classical data.  In particular, we can get \eqref{eq:trace} from a path integral over classical states on a coadjoint orbit.  For $SO(3)$, this path integral takes the form
\beq\label{eq:path}
\frac{1}{p} \int_{S_{J^2}^2} e^{i \epsilon J_z} \mu_J.
\eeq
The integral is over angular momentum vectors, $\vec{J} = (J_x,J_y,J_z) \in S^2_{J^2}$, on the coadjoint orbit.   The integration measure is $\mu_J = (2\pi)^{-1}d\varphi dJ_z$, $\varphi \in [0,2\pi]$.  The integral over $\varphi$ is trivial.  The remaining integral gives
\beq\label{eq:kirillov}
\frac{1}{i\epsilon p} (e^{i \epsilon |J|} - e^{-i \epsilon |J|}).
\eeq
The definition of the normalization factor, $p$, is a bit subtle \cite{kirillov2004lectures} but the result is $p=\frac{1}{i\epsilon}(e^{i\epsilon/2}-e^{-i\epsilon/2})$.  We now get the trace \eqref{eq:trace} from the path integral by making the identifications $q=e^{i\epsilon}$ and $|J| = \ell+1/2$.  

This formula for the trace improves the situation somewhat.  We gave a fairly explicit description of the coadjoint orbits of \sdiff\ in Section \ref{sec:labels}, so we might contemplate computing the traces of the quantum $\hat{\xi}_{\ell 0}$ operators using a generalized version of the path integral \eqref{eq:path}.  Unfortunately, we do not know the integration measure on \sdiff\ coadjoint orbits.

To go further, observe that the path integral localizes, in the following sense.  The classical version of the $\hat{J}_z$ operator is  the $J_z$ function on phase space.  $J_z$ generates an infinitesimal rotation of the coadjoint orbit.  This rotation has two fixed points, the north and south poles $(J_x,J_y,J_z) = (0,0,\pm |J|) \in S^2_{J^2} $.   The two terms in the evaluation of the path integral \eqref{eq:kirillov} come from these two fixed points.   This is an example of a very general phenomenon.  According to the Duistermaat-Heckman theorem \cite{duistermaat1982variation}, oscillatory integrals such as \eqref{eq:path} that arise from a $U(1)$ Hamiltonian action on a symplectic manifold always localize to a sum of terms coming from the fixed points of the $U(1)$ action.  The means we can bypass the need for an integration measure on the coadjoint orbit.  Instead, we can get the traces we are interested in directly from the fixed points of a $U(1)$ action on the orbit.   

In particular, we have the Atiyah-Bott fixed point formula \cite{atiyah1967lefschetz,atiyah1968lefschetz,witten1988coadjoint},
\beq\label{eq:ab}
\Tr q^{\hat{O}} =	\sum_{\substack{
					{\rm fixed}\\
					{\rm points}
                               }}
                               (-1)^s q^h \prod_k\frac{1}{1-q^{|n_k|}} .
\eeq
The lhs is the trace of a quantum operator, $\hat{O}$, in a fixed representation.  Suppose we have a classical realization of  the representation as a coadjoint orbit and the operator, $\hat{O}$, as a $U(1)$ transformation of the orbit.  Then the rhs of \eqref{eq:ab} gives the trace as a sum over terms defined at the fixed points of the $U(1)$ action.  Each term is weighted by $q^h$, where $h$ is the Hamiltonian generator of the $U(1)$ transformation. To define the $n_k$ and $s$ appearing in \eqref{eq:ab}, let $V$ be the Hamiltonian vector field on the coadjoint orbit that generates the $U(1)$ transformation.  Fixed points of the $U(1)$ action are the same thing as zeros of $V$.  Near a zero of $V$, expand
\beq\label{eq:complex}
iV = v -\bar{v}, \quad v = \sum_k n_k w_k \frac{\partial}{\partial w_k},
\eeq
where $w_k$ are local complex coordinates.  The $n_k$ so defined are integers and $s$ is the number of negative $n_k$.

Consider the $SO(3)$ trace \eqref{eq:trace} once more.  The operator $\hat{J}_z$ corresponds to  the classical function $J_z$ on phase space.  To get a vector field on $S^2_{J^2}$, we use the Kirillov-Kostant symplectic form,  $\omega_J = d\varphi \wedge dJ_z, \varphi = \arctan{(J_y/J_x)}$.  The vector field, $V$, defined by $\iota_V \omega_J= dJ_z$, is $V=\partial_\varphi$.  This is the Hamiltonian vector field generated by $J_z$.  It generates a $U(1)$ transformation of the orbit with fixed points at the poles, $(J_x,J_y,J_z) = (0,0,\pm |J|)$.  In coordinates,
\beq
V = J_y \frac{\partial}{\partial J_x} - J_x \frac{\partial}{\partial J_y}.
\eeq
To apply the fixed point formula, we need $V$ in complex coordinates \eqref{eq:complex}.   Our choice of complex coordinates is standard.  Separate charts are required for the northern and southern hemispheres.  Near the north pole $w=(J_x - i J_z)/(1+J_z)$ and near the south pole $w=(J_x+iJ_y)/(1-J_z)$.  We find
\begin{align}
iV = -w \frac{\partial}{\partial w} + \bar{w} \frac{\partial }{\partial \bar{w}} 	\quad	\mbox{(north pole)}, \quad
iV =w \frac{\partial}{\partial w} - \bar{w} \frac{\partial }{\partial \bar{w}} 	\quad	\mbox{(south pole)}.
\end{align}
Evidently $n=-1$ and $s=1$ at the north pole and $n=1$ and $s=0$ at the south pole.  

The Hamiltonian function generating the $U(1)$ action is defined up to a constant as $h=J_z+c$.  Evaluated at the fixed points, $h=\pm |J| + c$.  Now the Atiyah-Bott fixed point formula gives
\beq\label{eq:trace3}
\Tr q^{\hat{J}_z} = \frac{q^{|J|+c} - q^{-|J|+c}}{q-1}.
\eeq
We find a precise match with our two earlier computations of this trace upon identifying $|J| = \ell + 1/2$ and $c=1/2$.

\subsection{\sdiff}

We want to extend the proceeding discussion to \sdiff.  In particular, we want to use the Atiyah-Bott fixed point formula \eqref{eq:ab} to compute the traces of the $\hat{\xi}_{\ell0}$ operators.  

The classical $\xi_{\ell 0}$ functions are defined by \eqref{eq:coords}.  Evaluated on a one-form  $u\in {\rm SVect}(S^2)^*$,
\beq
\xi_{\ell 0}(u) = -\int_{S^2} Y_{\ell 0} du.
\eeq
Each of these functions generates a Hamiltonian flow on phase space by the usual formula,
\beq
\frac{d \xi_{\ell'm'}(t)}{dt} = \{ \xi_{\ell' m'},\xi_{\ell 0} \}   = n \times \nabla \{Y_{\ell 0},Y_{\ell' m'}\},
\eeq
where we used \eqref{eq:comm} and \eqref{eq:pbs} . The rhs is interpreted as a function on phase space using \eqref{eq:coords}. The flow is interpreted as follows.  We begin at $t=0$ at some $u_0\in {\rm SVect}(S^2)^*$.  The flow sends $u_0 \rightarrow u_t$ with coordinates $\xi_{\ell' m'}(t) \equiv \xi_{\ell' m'}(u_t)$.  

The flow generated by $\xi_{10}$ is
\beq\label{eq:flow}
\frac{d \xi_{\ell'm'}(t)}{dt} 
	= n \times \nabla \{Y_{10},Y_{\ell' m'}\}
	= i m' \xi_{\ell'm'},
\eeq
where we used \eqref{eq:basis} and \eqref{eq:Ypb}.  Here and below, we rescale $t$ to eliminate an overall numerical coefficient.  This flow has the explicit solution $\xi_{\ell'm'}(t)=e^{im't}\xi_{\ell'm'}(0)$.  It is periodic under $t\rightarrow t+2\pi$, so it generates a $U(1)$ action on coadjoint orbits.  We are therefore in a position to apply the Atiyah-Bott fixed point formula \eqref{eq:ab} to compute $\Tr q^{\hat{\xi}_{10}}$.

The flow generated by $\xi_{20}$ is
\beq
\frac{d \xi_{\ell'm'}(t)}{dt} 
	= n \times \nabla \{Y_{20},Y_{\ell' m'}\}
	= i m' \cos\theta \xi_{\ell'm'},
\eeq
which has the solution $\xi_{\ell'm'}(t)=e^{im'(\cos\theta)t}\xi_{\ell'm'}(0)$.  Now the flow is not periodic in $t$.  It is not hard to see that the flow is not periodic for all $\ell>1$.   So the Atiyah-Bott formula does not apply in these cases.   The remainder of this section is devoted to the case $\ell=1$.

\subsection{Fixed points}

To apply the Atiyah-Bott formula \eqref{eq:ab}, we need to fix a coadjoint orbit.  To fix a coadjoint orbit, we will pick an element $u\in{\rm SVect}(S^2)^*$ and set
\beq
\mathcal{O}_u = \{\eta \cdot u | \eta \in {\rm SDiff}(S^2)\}.
\eeq
In light of \eqref{eq:coords}, it is natural to define a collection of basis elements, $u_{\ell m}$, for \svectstar\ by 
\beq\label{eq:ubasis}
du_{\ell m} = -Y^*_{\ell m} \mu.
\eeq
On this basis, the coordinate functions \eqref{eq:coords} are simply $\xi_{\ell m}(u_{\ell' m'}) = \delta_{\ell \ell'}\delta_{mm'}.$  It is not too hard to solve for
\beq
u_{\ell m} = \frac{1}{\ell(\ell+1)}\left[(\sin\theta)^{-1} \partial_\varphi Y^*_{\ell m} d\theta - \sin\theta \partial_\theta Y^*_{\ell m} d\varphi\right].
\eeq
Define $\mathcal{O}_{\ell m} \equiv \mathcal{O}_{u_{\ell m}}$.  We will use the Atiyah-Bott formula to compute $\Tr q^{\hat{\xi}_{10}}$ on the coadjoint orbit $\mathcal{O}_{10}$. 

We need to find the fixed points of the flow generated by $\xi_{10}$ on $\mathcal{O}_{10}$.  Consider the action of the flow on $u_{10}\in \mathcal{O}_{10}$.  The only nonzero coordinate function is $\xi_{10}(u_{10}) = 1$, and this is fixed by \eqref{eq:flow}.  So $u_{10}$ is a fixed point.

$-u_{10}$ is also a fixed point.  The only thing to check is that it lies on the coadjoint orbit, $\mathcal{O}_{10}$.  This amounts to checking that $u_{10}$ and $-u_{10}$ have the same vorticity moments \eqref{eq:Ik}.  The vorticity functions are given by \eqref{eq:ubasis} as $\pm du_{10}/\mu = \mp Y_{10} = \mp (3/4\pi)^{1/2} \cos\theta$.  The odd vorticity moments, $I_{2n+1}$, vanish and the even vorticity moments, $I_{2n}$, of $\pm u_{10}$ are the same.  So $\pm u_{10}$ lie on the same coadjoint orbit.  We now have two fixed points of the flow generated by $\xi_{10}$ on $\mathcal{O}_{10}$.

More generally, any linear combination $u=\sum_\ell a_\ell u_{\ell 0}$ in \svectstar\ is fixed by the flow \eqref{eq:flow}.  The challenge is finding solutions on $\mathcal{O}_{10}$.  We have been unable to find exact solutions other than $\pm u_{10}$.  We will compute the contributions of these two fixed points to $\Tr q^{\hat{\xi}_{10}}$.  This will not give the full answer, but it will give some insight into the trace.

\subsection{Complex coordinates}

The flow generated by $\xi_{10}$ arises from a Hamiltonian vector field, $V_1$, defined by
\beq
V_1(F) \equiv \{F,\xi_{10} \},
\eeq
where $F$ is an arbitrary function on \svectstar.  The components of this vector field are $V_1(\xi_{\ell m}) = \{\xi_{\ell m},\xi_{10}\} =  i m \xi_{\ell m}$.   So
\beq
V_1 = \sum_{\ell =1}^\infty \sum_{m=-\ell}^\ell im \xi_{\ell m} \frac{\partial}{\partial \xi_{\ell m}}.
\eeq
$V_1$ can be decomposed as in \eqref{eq:complex} by setting $\overline{\xi}_{\ell m} \equiv \xi_{\ell,-m}$, which gives
\beq
iV_1 = v_1 - \bar{v}_1, \quad v_1 =  -\sum_{\ell=1}^\infty \sum_{m=1}^\ell m \xi_{\ell m}\frac{\partial}{\partial \xi_{\ell m}}.
\eeq

\subsection{Result and discussion}

We are now ready to apply the Atiyah-Bott fixed point formula \eqref{eq:ab}.  It gives the formal expression
\beq\label{eq:tracexi}
\Tr q^{\hat{\xi}_{10}} = (-1)^s q^{1+c} \prod_{\ell=1}^\infty \prod_{m=1}^\ell \frac{1}{1-q^m}
				+ q^{-1+c} \prod_{\ell=1}^\infty \prod_{m=1}^\ell \frac{1}{1-q^m}
				+\dots
\eeq
The first term is the contribution from $u_{10}$ and the second term is the contribution from $-u_{10}$.  The dots indicate contributions from additional fixed points.  Up to an overall constant $c$, the Hamiltonian function generating the $U(1)$ transformation is $\xi_{10}(\pm u_{10}) = \pm 1$ at the leading two fixed points.  This provides the factors of $q^{1+c}$ and $q^{-1+c}$ multiplying the first two terms.

Unfortunately, the trace \eqref{eq:tracexi} is divergent.  The infinite product, $\prod_\ell \prod_m \frac{1}{1-q^m}$, does not have a sensible $q$-series expansion analogous to the middle expression in \eqref{eq:trace}. The factor of $(-1)^s$ multiplying the leading term is also ill-defined.  It is formally equal to $(-1)^\infty$.

Evidently, some regularization is required.  A crude but straightforward possibility is to cut off the infinite product at some $\ell=\ell_{\rm max}$.  For example, take $\ell_{\rm max}=1$.  In this case, the trace \eqref{eq:tracexi} is simply
\beq\label{eq:tracexi2}
\frac{q^{1+c}-q^{-1+c}}{q-1}.
\eeq
In this truncation, there are no fixed points beyond $\pm u_{10}$ and \eqref{eq:tracexi2} is exact.  It matches the $SO(3)$ trace \eqref{eq:trace3} with $|J|=1$ and $c=1/2$.  Actually, we get exactly \eqref{eq:trace3} by repeating our calculations on the \sdiff\ coadjoint orbit passing through $\pm|J| u_{10}$.  Now the coordinate functions of the fixed points are $\xi_{10}(\pm |J|u_{10}) = \pm |J|$.  The truncation of the trace with $\ell_{\rm max} = 1$ becomes
\beq
\frac{q^{|J|+c}-q^{-|J|+c}}{q-1},
\eeq
which precisely matches \eqref{eq:trace3}.

The fact that we can recover the $SO(3)$ trace from a truncation of \eqref{eq:tracexi} is perhaps not so surprising because $SO(3)$ is a subgroup of \sdiff.  Indeed, after suitable rescalings, the commutators of the $\ell=1$ modes of \svect\ are
\beq
[ \xi_{1,1}, \xi_{1,-1} ] = 2\xi_{1,0}, \quad
[ \xi_{1,0}, \xi_{1,1} ] = \xi_{1,1}, \quad
[ \xi_{1,0}, \xi_{1,-1} ] = -\xi_{1,-1},
\eeq
which is the same thing as the $\mathfrak{so}(3)$ Lie algebra \eqref{eq:liealg} in the basis $\hat{J}_z,\hat{J}_\pm \equiv \hat{J}_x \pm i \hat{J}_y$,
\beq
[ \hat{J}_+, \hat{J}_- ] = 2\hat{J}_z, \quad
[ \hat{J}_z, \hat{J}_+ ] = \hat{J}_+, \quad
[ \hat{J}_z, \hat{J}_- ] = -\hat{J}_-.
\eeq

This example points to one interpretation of the divergent series \eqref{eq:tracexi}.  The trace itself is divergent, but it has well-defined truncations which are related to the structure of \sdiff.  The relationship to $SO(3)$ is fairly simple.  To get more intricate examples, recall that \sdiff\ is related to the  large $N$ limit of $SU(N)$ \cite{fairlie1989trigonometric,hoppe1989diffeomorphism,bordemann1991gl}.  It would be interesting to study the relationship between truncations of the trace with $\ell_{\rm max}>1$ and traces of $SU(N)$ operators with $N>2$.  Exploring this structure will require understanding the contributions from fixed points other than $\pm u_{10}$ to the trace \eqref{eq:tracexi}.

\section{Discussion}
\label{sec:discuss}

This work was originally motivated by the recent appearance of \sdiff\ as a subgroup of two newly defined and closely related symmetry groups in general relativity.  One of these groups acts on the event horizons of four dimensional black holes \cite{Donnay:2015abr,Donnay:2016ejv,Chandrasekaran:2018aop} and the other acts on the asymptotic boundaries of four dimensional asymptotically flat spacetimes \cite{Campiglia:2015yka,Compere:2018ylh,Flanagan:2019vbl}.   The full symmetry group that appears in these examples is ${\rm Diff}(S^2)\ltimes C^\infty(S^2)$.  Incidentally, the latter group is related to the symmetry group of the compressible Euler equations for fluid flow on the two-sphere \cite{marsden1984semidirect}.  This might not be a coincidence: the dynamics of null surfaces in general relativity can be described as a kind of compressible hydrodynamics \cite{thorne1986black,Penna:2015gza,Penna:2017vms}.

Part of the recent interest in these symmetries comes from experience with asymptotic symmetry groups of three dimensional gravity.  For a certain choice of boundary conditions, the asymptotic symmetry group of three dimensional gravity is either the two dimensional conformal group \cite{brown1986central} (if the cosmological constant is nonzero) or the ${\rm BMS}_3$ group \cite{barnich2007classical} (if the cosmological constant vanishes).  These groups each have a rich representation theory and a lot has been learned about three dimensional gravity using this representation theory (for example, see \cite{Barnich:2014kra,Oblak:2015sea,Barnich:2015uva,Garbarz:2015lua,Oblak:2016eij} for some recent work on the asymptotically flat case).  It would be very interesting if ${\rm Diff}(S^2)\ltimes C^\infty(S^2)$ symmetry turns out to play a similar role in four dimensions.

The attempts to get ${\rm Diff}(S^2)\ltimes C^\infty(S^2)$  to act in four dimensions are not without problems.
At the event horizon, the action of the symmetry appears to be somewhat observer dependent.  At asymptotic infinity, it is difficult to get finite ${\rm Diff}(S^2)\ltimes C^\infty(S^2)$  Noether charges.  There are ongoing attempts to solve these problems.

Our strategy for the present paper has been to suppose these problems can be solved and ask whether the group has a good representation theory.  We focused on \sdiff\ for simplicity but we expect general lessons learned here also apply to ${\rm Diff}(S^2)\ltimes C^\infty(S^2)$ (and higher dimensional generalizations).  Our main result is formula \eqref{eq:tracexi} for $\Tr q^{\hat{\xi}_{10}}$.  It is divergent, although we have shown that it has well-defined truncations related to the structure of \sdiff.  A pessimistic interpretation of the divergence is that \sdiff\ does not have an interesting representation theory.  This would suggest that even if ${\rm Diff}(S^2)\ltimes C^\infty(S^2)$ is a symmetry of four dimensional general relativity, the symmetries can only be relevant for classical physics.  An optimistic interpretation is that \sdiff\ might have a good representation theory but a new idea is needed to make sense of it.  A perhaps intermediate interpretation is that \sdiff\ can only be realized in quantum field theory as an approximate symmetry, something that emerges, for example, in the large $N$ limit of an underlying $SU(N)$ symmetry.

\acknowledgments

We are grateful to Blagoje Oblak for discussions and for comments on an earlier version of the manuscript.

\bibliographystyle{JHEP}
\bibliography{ms}

\end{document}